\documentclass[11pt,prd,preprint,nofootinbib]{revtex4}

\usepackage{amsmath}
\usepackage{amsfonts}
\usepackage[colorlinks,hyperindex]{hyperref}

\mathchardef\mhyphen="2D

\newcommand{\nc}{\newcommand}
\newcommand\fft[2]{\frac{#1}{#2}}
\newcommand\ft[2]{{\textstyle\frac{#1}{#2}}}
\newcommand\nn{{\nonumber}}
\newcommand{\beq}{\begin{equation}}
\newcommand{\eq}{\end{equation}}

\nc{\bea}{\begin{eqnarray}} \nc{\ea}{\end{eqnarray}} \nc{\be}{\begin{equation}} \nc{\ee}{\end{equation}} \nc{\barr}{\begin{array}}
\nc{\earr}{\end{array}}

\nc{\tr}{\text{tr}\,}

\makeindex

\begin{document}
\preprint{MCTP-14-23}

\title{$c-a$ from the $\mathcal N=1$ superconformal index}

\author{Arash Arabi Ardehali}
\email{ardehali@umich.edu}
\author{James T. Liu}
\email{jimliu@umich.edu}
\affiliation{Michigan Center for Theoretical Physics, Randall Laboratory of Physics,\\
The University of Michigan, Ann Arbor, MI 48109--1040, USA}
\author{Phillip Szepietowski}
\email{pgs8b@virginia.edu}
\affiliation{Department of Physics, University of Virginia,\\
Box 400714, Charlottesville, VA 22904, USA}

\begin{abstract}
We present a prescription for obtaining the difference of the
central charges, $c-a$, of a four dimensional superconformal quantum
field theory from its single-trace index. The formula is derived
from a one-loop holographic computation, but is expected to be valid
independent of holography. We demonstrate the prescription with
several holographic and non-holographic examples. As an application
of our formula, we show the AdS/CFT matching of $c-a$ for arbitrary
toric quiver CFTs without adjoint matter that are dual to smooth
Sasaki-Einstein 5-manifolds.
\end{abstract}


\maketitle

\section{Introduction}

Over the last few decades, there has been enormous progress in the understanding
of supersymmetric field theories, both from the field theory side and from AdS/CFT.
Of particular interest are superconformal gauge theories admitting AdS duals.  As is
often the case when exploring strong/weak coupling dualities,
AdS/CFT is most powerful when applied to BPS sectors that are protected
from quantum corrections and therefore can be understood at all
couplings. In this context, the $\mathcal N=1$ superconformal index
\cite{Romelsberger:2005eg,Kinney:2005ej} has received much attention
as a quantity which efficiently encodes the protected information in
superconformal quantum field theories.

The superconformal index is defined as a refined
Witten index for the theory in radial quantization. In four
dimensions it is given by
\begin{equation}
\mathcal{I}^R_{s.t.}
(t,y;a_i)=\mathrm{Tr}_{s.t.}(-1)^{F}e^{-\beta\delta}t^{-2(E+j_{2})/3}
y^{2j_{1}}\prod a_i^{2s_i},
\label{eq:IRst}
\end{equation}
where $\delta=E-\fft32 r-2j_{2}$, and $\{E,j_1 ,j_2 ,r;s_i\}$ are
the quantum numbers of the superconformal group SU(2,2$|$1) and the
global flavor symmetries, and $\{t,y;a_i\}$ are the corresponding
fugacities%
\footnote{As we discuss below, we have chosen the
normalization of the exponent of $t$ such that a chiral primary
operator would contribute with an exponent given by the negative of its
R-charge.  Hence our parameter $t$ is the inverse of the one in
\cite{Romelsberger:2007,Dolan:2008,Spiridonov:2011,Kutasov:2014},
and the inverse cubed of the one in \cite{Kinney:2005ej,Gadde:2010en}.}.
As defined this is the right-handed index. One can also
define a left-handed index $\mathcal I_{s.t.}^L$ in which one
replaces $r$ with $-r$ and swaps $j_1$ and $j_2$ in both the
definition of the index and of $\delta$.
Only states with $\delta = 0$ contribute to the index; thus it is
independent of $\beta$. This condition means that only states which
lie within shortened representations of the superconformal algebra
will contribute to the index. The index is hence a protected
quantity and is independent of the coupling, and therefore can be
employed to test various supersymmetric weak/strong dualities
\cite{Kinney:2005ej,Romelsberger:2007,Dolan:2008,Spiridonov:2011,Kutasov:2014,Gadde:2010en}.

In this article we demonstrate that it is possible to obtain the difference of central
charges, $c-a$, from the large-$N$ single-trace superconformal index.  This difference
is related to the mixed U(1)$_R$ chiral anomaly \cite{Anselmi:1997} via $c-a =
-\fft1{16}\mathrm{Tr} R$, where the trace is over the fundamental fermions that can run
in the loop of the associated triangle diagram.   Thus there is a direct connection between
the superconformal index and $\mathrm{Tr} R$.  The combination $c-a$ is
known to partially determine the subleading logarithmic contribution
of quantum fields to the entropy of various extremal and non-extremal black-hole backgrounds
\cite{Sen:2012,Sen:2013,Keeler:2014}. It also appears in the
subleading correction to the shear viscosity-to-entropy ratio in gauge theory plasmas admitting
a holographic dual \cite{Kats:2009}.

The expression we have obtained for extracting $c-a$ from the large-$N$ single-trace index is
\begin{eqnarray}\label{eq:c-aIndex}
c-a &= &\lim_{t\to1}-\frac{1}{32}\left(t\partial_t
+1\right)\left(6(y\partial_y)^2-1\right)\nn\\
&&\times\left[(1-t^{-1} y)(1-t^{-1} y^{-1})\mathcal
I^+_{s.t.}(t,y;a_i)\right]
\Big|^{\mbox{\scriptsize{finite}}}_{y=1,a_i=1},
\end{eqnarray}
where $\mathcal I^+_{s.t.} \equiv \frac{1}{2}(\mathcal I^R_{s.t.}+
\mathcal I^L_{s.t.})$, and the fugacities are set to one after
acting with the differential operator on the index. Note that the
factor $(1-t^{-1} y)(1-t^{-1} y^{-1})$ multiplying the single-trace
index removes the contribution from descendant states. The result
obtained is often divergent, as we are working in the large-$N$
limit, so the prescription is that the finite term in an expansion
about $t=1$ yields the value of $c-a$.

Some care must be
taken in removing the divergent terms in (\ref{eq:c-aIndex}), as different regularization
procedures will affect the finite term and thus the result for $c-a$.  For example, a
double pole of the form $1/(t-1)^2$ will be converted into $t^2/(t-1)^2=1/(t-1)^2+2/(t-1)
+1$ under the replacement $t\to t^{-1}$.  What we find is that the choice of $t$ in (\ref{eq:IRst})
is what yields $c-a$ after simply dropping the pole terms
in the Laurent expansion around $t=1$. Moreover, as we will see in later sections, this choice is
unique in that it yields only a double pole (and no simple pole) in $t$ for all of the large-$N$
examples we consider.
Of course, if the divergent terms were absent, then the result for $c-a$ would be manifestly
independent of the normalization of the exponent of $t$. At finite-$N$ this is in fact the case, {\it i.e.} there is no divergence and $c-a$ is independent of the normalization. However, in the infinite-$N$ limit this divergence arises due to the infinite sums encountered in this limit.

For a holographic CFT, Eq.~(\ref{eq:c-aIndex}) can be thought of as an
expression that allows one to obtain $\mathrm{Tr} R$ (or $c-a$) from
the protected mesonic single-trace states of the CFT at large 't~Hooft coupling.
In fact, we came upon this expression starting from
a one-loop holographic Weyl anomaly computation which amounts to
\cite{Mansfield:2000,Ardehali:2013xya}
\begin{equation}
c-a=-\fft1{360}\sum(-1)^F(E_0-2)d(j_1,j_2)\left(1+f(j_1)+f(j_2)\right).
\label{eq:c-aexpr}
\end{equation}
Here the sum is over all dual supergravity fields with AdS$_5$
quantum numbers $(E_0,j_1,j_2;r)$, and where $d(j_1,j_2) =
(2j_1+1)(2j_2+1)$ and $f(j) = j(j+1)[6j(j+1)-7].$  It is easy to see that
only shortened multiplets of SU$(2,2|1)$ contribute to this quantity
\cite{Ardehali:2013xya}, and that it has the characteristics of a combination
of left and right indices.

The shortened multiplets of SU$(2,2|1)$ are listed in
Table~\ref{tbl:IndexContributions}, along with their contributions to the
single-trace index.  The chiral and SLII multiplets contribute to the
right-handed index, while the CP-conjugate multiplets, namely the
anti-chiral and SLI multiplets, contribute to the
left-handed index.  Conserved multiplets, which are CP self-conjugate, contribute
to both.

\begin{table*}[t]
\centering
\begin{tabular}{|l|l|l|c|}
\hline
Shortening&Condition&Representation& $(1-t^{-1} y)(1-t^{-1} y^{-1})\mathcal{I}^+$\\
\hline
conserved&$E_0=2+j_1+j_2$, $\fft32r=j_1-j_2$&$\mathcal D(E_0,j_1,j_2,r)$&$ \ft12(-1)^{2(j_1+j_2)+1} t^{-(2E_0+2j_2+2)/3}\chi_{j_1}(y) +(j_1\leftrightarrow j_2)$\\
\hline
chiral&$E_0=\fft32r$&$\mathcal D(E_0,j_1,0,r)$&$ \ft12(-1)^{2j_1} t^{-2E_0/3}\chi_{j_1}(y)$\\
\hline
anti-chiral&$E_0=-\fft32r$&$\mathcal D(E_0,0,j_2,r)$&$ \ft12(-1)^{2j_2} t^{-2E_0/3}\chi_{j_2}(y)$\\
\hline
SLI&$E_0=2+2j_1-\fft32r$&$\mathcal D(E_0,j_1,j_2,r)$&$\ft12(-1)^{2(j_1+j_2)+1} t^{-(2E_0+2j_1+2)/3}\chi_{j_2}(y)$\\
\hline
SLII&$E_0=2+2j_2+\fft32r$&$\mathcal D(E_0,j_1,j_2,r)$&$\ft12(-1)^{2(j_1+j_2)+1} t^{-(2E_0+2j_2+2)/3}\chi_{j_1}(y)$\\
\hline
\end{tabular}
\caption{Contributions to the superconformal index from the various
shortened multiplets.\label{tbl:IndexContributions}}
\end{table*}

In order to relate $c-a$ to the index, we first consider the chiral and SLII multiplets.
The contribution to $c-a$ from a generic chiral multiplet $\mathcal D(E_0,j_1,0;r)$
is given by summing (\ref{eq:c-aexpr}) over all the fields of the multiplet. This yields
\begin{equation}
(c-a)\big|_{\text{chiral}}=
-\fft1{192}(-1)^{2j_1}(2E_0-3)(2j_1+1)
\left(1-8j_1(j_1+1)\right).
\label{eq:CHc-a}
\end{equation}
Similarly, a generic SLII multiplet $\mathcal D(E_0,j_1,j_2;r)$ contributes
\begin{equation}
(c-a)\big|_{\text{SLII}}=\fft1{192}(-1)^{2j_1+2j_2}(2E_0+2j_2-1)(2j_1+1)
\left(1-8j_1(j_1+1)\right).\label{eq:sl2c-a}
\end{equation}
It is now possible to see how these expressions may be obtained from the contributions to
the right-handed index given in Table~\ref{tbl:IndexContributions}.
Since the SU(2) character $\chi_j(y)$ is given by
\begin{equation}
\chi_j(y)=\fft{y^{2j+1}-y^{-(2j+1)}}{y-y^{-1}},
\end{equation}
the differential operator $(6(y\partial_y)^2-1)$ acting on the
contributions to the index gives $(2j+1)[8j(j+1)-1]$ when $y$ is set
to one.  The operator $(t\partial_t+1)$ then produces the
$E_0$-dependent factors in (\ref{eq:CHc-a}) and (\ref{eq:sl2c-a}).

The CP conjugate multiplets (anti-chiral and SLI) contribute similarly to
(\ref{eq:CHc-a}) and (\ref{eq:sl2c-a}) with the appropriate replacement of quantum numbers, and
are accounted for in the left-handed index.  Finally, since conserved multiplets
contribute as the sum of one SLI and one SLII multiplet, they are implicitly included in both
the left- and right-handed indices.  Our key observation is that the contribution to $c-a$ has
an uniform expression for every single bulk multiplet.  Hence a single differential operator acting
on the index can yield the appropriate contribution to $c-a$ regardless of the shortening condition.
Summing over all multiplets, one finally arrives at (\ref{eq:c-aIndex}), where the
index is now the single-particle supergravity index which is equal
to the single-trace index of the SCFT. As a side comment, note that the index thus provides a natural regulator for the Kaluza-Klein sums encountered in the holographic $c-a$ calculations of \cite{Ardehali:2013gra,Ardehali:2013xla,Ardehali:2013xya}.

Although Eq.~(\ref{eq:c-aIndex}) was derived via a holographic
calculation, it refers only to field
theoretic objects, and so we conjecture it to be true in general for
all superconformal field theories, regardless of whether they admit
a dual holographic description or not.  This is most straightforward in
the large-$N$ limit, where the single-trace index is well-defined.
However, Eq.~(\ref{eq:c-aIndex}) can be extended away from the
large-$N$ limit by simply replacing the single-trace index by the
plethystic log of the full index \cite{Hanany:2006,Hanany:2007}.
(Another connection between the central charges and the
full index is suggested by the results of \cite{Assel:2014}, which
demonstrated that $c$ and $a$ are independently encoded in the
prefactor relating the supersymmetric partition function on $S^3\times S^1$
to the index.)
In the remainder of this article we discuss the application of
(\ref{eq:c-aIndex}) to a variety of examples.

\section{Examples}

Since all the theories we consider are CP invariant, we have
$\mathcal I^+_{s.t.} = \mathcal I^R_{s.t.}= \mathcal I^L_{s.t.}$,
and therefore we dispense with the $+$ superscript of $\mathcal
I_{s.t.}$ in the following.

\subsection{Holographic Theories}

We begin with an examination of Eq.~(\ref{eq:c-aIndex}) in the
context of four-dimensional superconformal field theories admitting
an AdS$_5$ dual at large $N$. For such theories it is well-known
that even though $c$ and $a$ are both of order $N^2$, their
difference $c-a$ is $\mathcal{O}(1)$, and therefore it vanishes at
leading order \cite{Henningson:1998gx,Benvenuti:2004gx}. Our
prescription thus gives an $\mathcal{O}(1/N^2)$ correction to the
leading order quantities at large $N$.

\subsubsection{The $\mathcal{N}=4$ SYM theory and the $\mathbb Z_2$ orbifold}

We first consider $\mathcal N=4$ SYM theory dual to IIB string theory on AdS$_5\times S^5$.
The $\mathcal N=1$ large-$N$ single-trace index for $\mathcal{N}=4$ SYM
with gauge group SU($N$) is \cite{Kinney:2005ej}
\begin{equation}
\mathcal{I}_{s.t.}=\frac{3}{t^{2/3}-1} - \frac{3t^{-2/3}
(1-t^{-2/3})}{(1-t^{-1} y)(1-t^{-1} /y)}.
\end{equation}
Inserting this into (\ref{eq:c-aIndex}) gives
\begin{equation}\label{eq:N=4c-a}
c-a=\left[-\frac{27}{16(t-1)^2}+\mathcal{O}(t-1)\right]^{\mbox{\scriptsize{finite}}}_{t\rightarrow
1}=0,
\end{equation}
as expected, since $c=a$ exactly for the $\mathcal N=4$ theory.  It
is also possible to verify that the decoupled U(1) has vanishing
contribution to $c-a$, so the result (\ref{eq:N=4c-a}) holds for the
U$(N)$ case as well. While this example is somewhat trivial, it
nevertheless demonstrates that a divergent term needs to be removed
when extracting $c-a$ from the single-trace index.  Moreover, a
possible simple pole divergence is absent, but would have been
present had we not made the choice for $t$ given in (\ref{eq:IRst}).%
\footnote{The vanishing of the simple pole holds for all the examples we have investigated.
This appears to be a universal property of the Laurent expansion of (\ref{eq:c-aIndex}),
although it is sensitive to the normalization of the exponent of $t$ in (\ref{eq:IRst}). It would be interesting to understand this further.}
The double pole divergence appears to be an artifact of the
large-$N$ limit, and we will comment further on this below.

We now consider the $S^5/\mathbb Z_2$ orbifold, which
has $\mathcal{N}=2$ supersymmetry. The index was computed in
\cite{Nakayama:2005,Gadde:2009dj} and is given by
\begin{equation}
\mathcal{I}_{s.t.}=\frac{2}{t^{2/3}-1}+\frac{2}{t^{4/3}-1} -
\frac{2t^{-2/3} (1-t^{-2/3})}{(1-t^{-1} y)(1-t^{-1}/y)}.
\label{eq:Z2orbi}
\end{equation}
Using this expression in (\ref{eq:c-aIndex}) gives
\begin{equation}\label{eq:N=2c-a}
c-a=\left[-\frac{27}{16(t-1)^2}+\fft1{12}+\mathcal{O}(t-1)\right]^{\mbox{\scriptsize{finite}}}_{t\rightarrow
1}=\fft1{12},
\end{equation}
which matches the field theory result.

The computation can be extended to more general abelian $\mathcal
N=2$ orbifolds of $S^5$, although we have not explicitly done so
here.  Instead we turn next to toric quivers without adjoint matter
and with smooth Sasaki-Einstein dual geometries, these encompass $\mathcal
N=1$ orbifolds of $S^5$ as a special case. To be precise, this is true of the
$S^5/\mathbb{Z}_n$ orbifold theories only for odd $n$. The even $n$
orbifolds are not special cases of the toric theories considered
below because their dual geometry is singular. However, their index
and value of $c-a$ are obtained correctly if we treat them as
limiting cases of the toric $Y^{p,q}$ theories
\cite{Ardehali:2013xla}.

\subsubsection{Toric quivers without adjoint matter}

We now apply the relation (\ref{eq:c-aIndex}) to the class of toric
quiver gauge theories without adjoint matter dual to smooth
Sasaki-Einstein five-manifolds. These include the theories dual to
$\mathcal N=1$ orbifolds of $S^5$ as well as the $Y^{p,q}$ manifolds
which were discussed in
\cite{Ardehali:2013gra,Ardehali:2013xya,Ardehali:2013xla}. Here we
reproduce the $c-a$ results of those papers utilizing the relation
to the superconformal index.

The index for such a toric gauge theory was reported in
\cite{Eager:2012hx}. It is given by
\begin{equation}
\mathcal{I}_{s.t.}=\sum_i
\frac{1}{t^{r_i/3}-1},\label{eq:toricIndex}
\end{equation}
where $r_i$ are the $R$-charges of extremal BPS mesons, to be
determined by $a$-maximization \cite{Intriligator:2003jj}. In fact, $\sum r_i$ can be seen from
comparing equations (3.13) and (3.17) in \cite{Eager:2012hx} to be
given by
\begin{equation}
\sum r_i=6(\mbox{\# nodes in the quiver}).\label{eq:sumRule}
\end{equation}
Note that in obtaining (\ref{eq:sumRule}) one must assume $0\le
R_i\le 2$, where $R_i$ are the $R$-charges of chiral fields in the
quiver; the lower bound comes from unitarity while the upper bound
follows from the fact that in toric quiver theories all chiral
fields participate in superpotential terms \cite{Franco:2006} that
have $R$-charge equal to 2.

Applying (\ref{eq:c-aIndex}) to (\ref{eq:toricIndex}) gives
\begin{equation}
c-a=\left[\frac{1}{96}\sum_i
\left(3(1-t^{-2})\frac{1}{t^{r_i/3}-1}-(1+10t^{-1}+t^{-2})\frac{r_i
t^{-r_i/3}}{(1-t^{-r_i/3})^2}\right)\right]^{\mbox{\scriptsize{finite}}}_{t\rightarrow
1},
\end{equation}
which upon expanding around $t=1$ yields
\begin{equation}
c-a = \left[-\frac{9}{8(t-1)^2}\sum_{i=1}^{n_z} \frac{1}{r_i} +
\frac{1}{96}\sum_{i=1}^{n_z}
r_i\right]^{\mbox{\scriptsize{finite}}}_{t\rightarrow 1} =
\frac{1}{96}\sum_{i=1}^{n_z} r_i.\label{eq:polToric}
\end{equation}
Comparing (\ref{eq:polToric}) with (\ref{eq:sumRule}) then yields
\begin{equation}
c-a=\fft1{16}(\mbox{\# nodes in the quiver}).\label{eq:c-aToric}
\end{equation}
This matches the expected result for $c-a$ based on the decoupling of
a U(1) at each node in the quiver.
(Since there are no adjoints in the quiver, there are no additional
$\mathcal{O}(1)$ contributions to $c-a$ in the field theoretical
computation through $c-a=-\fft1{16}\mathrm{Tr}R$.)

Note that since Eq.~(\ref{eq:c-aIndex}) is derived from a
one-loop holographic computation, and since the AdS/CFT matching of
the index has been demonstrated in \cite{Eager:2012hx} for the theories
described above, the success of (\ref{eq:c-aIndex}) for the theories
considered so far may be interpreted as a successful AdS/CFT
matching of $c-a$. This significantly generalizes our earlier
results in
\cite{Ardehali:2013gra,Ardehali:2013xya,Ardehali:2013xla}.


\subsubsection{Suspended Pinch Point}

Next, we will illustrate the prescription (\ref{eq:c-aIndex}) for a
theory with a singular holographic dual geometry, namely, the
Suspended Pinch Point (SPP) quiver CFT \cite{Morrison:1998}.
This is a toric gauge theory with three SU$(N)$ gauge factors, one chiral
multiplet with $R$-charge $2-2/\sqrt{3}$ in the adjoint of one of
the gauge factors and a number of chiral bifundamental fields
\cite{Franco:2006}. Since, in the absence of fundamental matter,
$\mathrm{Tr} R$ for any holographic CFT is
order one (contrary to $a$ or $c$ which are of order $N^2$), we may
count only the order one contributions to it and be sure that order
$N^2$ contributions cancel. $\mathrm{Tr} R$ in this case is
$-3-(1-2/\sqrt{3})$, where the first term comes from the three
gauginos, and the second term from the fermion in the adjoint
chiral. Thus $c-a=-\fft1{16}\mathrm{Tr} R = (6-\sqrt{3})/24.$

We now demonstrate that this result can be reproduced via Eq.~(\ref{eq:c-aIndex}).
We need the large-$N$ index of the SPP quiver,
which can be computed following the approach of \cite{Gadde:2010en}.
The essential ingredient in the computation is the single-letter
index matrix $i_{SPP}$ for the quiver, which we find to be
\begin{widetext}
\begin{equation}\label{eq:iMat}\small
i_{SPP}=
\begin{pmatrix}
  i_V+(a_1a_2)^{-1}i_{\chi(2-2/\sqrt{3})}+(a_1a_2)i_{\bar{\chi}(2-2/\sqrt{3})} & a_2i_{\chi(1/\sqrt{3})}+a_1^{-1}i_{\bar{\chi}(1/\sqrt{3})} & a_1i_{\chi(1/\sqrt{3})}+a_2^{-1}i_{\bar{\chi}(1/\sqrt{3})}  \\
  a_1i_{\chi(1/\sqrt{3})}+a_2^{-1}i_{\bar{\chi}(1/\sqrt{3})} & i_V & a_2^{-1}i_{\chi(1-1/\sqrt{3})}+a_1i_{\bar{\chi}(1-1/\sqrt{3})} \\
  a_2i_{\chi(1/\sqrt{3})}+a_1^{-1}i_{\bar{\chi}(1/\sqrt{3})} & a_1^{-1}i_{\chi(1-1/\sqrt{3})}+a_2i_{\bar{\chi}(1-1/\sqrt{3})} &
  i_V
\end{pmatrix}.
\end{equation}
%
Following \cite{Gadde:2010en}, we are denoting by $i_V$,
$i_{\chi(r)}$, and $i_{\bar{\chi}(r)}$, respectively the
single-letter index of a vector multiplet, a chiral multiplet, and
an anti-chiral multiplet. We have introduced fugacities $a_1$ and
$a_2$ for the global U(1)$\times$U(1) flavor symmetry of the theory.
From the single-letter index matrix the large-$N$ single-trace index
of the theory can be computed as explained in \cite{Gadde:2010en}.
The result is
\begin{eqnarray}\label{eq:SPPindex}
\mathcal{I}_{s.t.}
(t,y;a_1,a_2)&=&-3+\frac{1}{1-(a_1a_2)^{-1}t^{-2/\sqrt{3}}}
+\frac{1}{1-a_1^{-1}t^{-1-1/\sqrt{3}}}
+\frac{1}{1-a_2^{-1}t^{-1-1/\sqrt{3}}}\nonumber\\
&&-\frac{2}{1-(a_1a_2)^{-1}t^{2-2/\sqrt{3}}}
+\frac{(a_1a_2)^{-1}t^{-2/\sqrt{3}}-a_1a_2t^{-2+2/\sqrt{3}}}{(1-t^{-1}
y)(1-t^{-1} y^{-1})}.
\end{eqnarray}
This could also be obtained from the results of \cite{Agarwal:2013}.
\end{widetext}

Setting $a_1$ and $a_2$ equal to one, and plugging the formula for the
index of the SPP quiver into (\ref{eq:c-aIndex}) we obtain
\begin{equation}\label{eq:SPPc-a}
c-a=\left[-\frac{27}{16(t-1)^2}+\frac{6-\sqrt{3}}{24}+\mathcal{O}(t-1)\right]^{\mbox{\scriptsize{finite}}}_{t\rightarrow
1}=\frac{6-\sqrt{3}}{24},
\end{equation}
which matches the field theory result. Note that the index (\ref{eq:SPPindex}) was obtained by
field theoretic means.  Thus, in particular, it is not necessary to work with a holographic dual
when applying the expression (\ref{eq:c-aIndex}) for $c-a$.

\subsubsection{del Pezzo theories}

For $k>3$, the quiver theories dual to dP$_k$ surfaces are not
toric, so they can not be considered as special cases of the
theories studied above. However, our prescription works very simply.
The index is \cite{Eager:2012hx}
\begin{equation}
\mathcal{I}_{s.t.}=(k+3) \frac{1}{t^2-1},
\end{equation}
which when combined with Eq.~(\ref{eq:c-aIndex}) gives
\begin{equation}\label{eq:N=dPc-a}
c-a=\left[-\frac{9(k+3)}{16(t-1)^2}+\fft{k+3}{16}+\mathcal{O}(t-1)\right]^{\mbox{\scriptsize{finite}}}_{y=1,t\rightarrow
1}=\fft{k+3}{16}.
\end{equation}
This matches the field theory result since the quivers have $k+3$
SU($N$) nodes and no adjoints.

\subsection{Non-holographic Theories}

We now consider several examples where the field theory does not
admit an AdS dual, and show that the expression (\ref{eq:c-aIndex})
continues to hold. Note, however, that we continue to work in a
large-$N$ limit where the single-trace index is well defined.

\subsubsection{U(1)$^N$ gauge theory with
matter}\label{subsubsec:U(1)N}

Consider a U(1)$^N$ gauge theory with $N_{\chi}$ chiral multiplets
(along with their conjugates) having $R$-charges $R_i$, and neutral
under the gauge group. We are not claiming such a theory exists as an
SCFT for generic values of $N,N_{\chi},$ and $R_i,$ however as described below
there exist specific values for which these theories do describe
particular SCFTs. With this example we only want to show that our
prescription is able to extract $-\fft{1}{16}\text{Tr}R$ from the
single-trace index even for very simple non-holographic systems. Note that
$(N,N_{\chi})=(0,1),$ and $(1,0)$ correspond to a single chiral
multiplet and a single U(1) vector multiplet, respectively. Also, for
the case $(N,N_{\chi})=(0,(N_c+1)^2+2(N_c+1)),$ with $2(N_c+1)$ of
the chiral multiplets having $R$-charge $\frac{N_c}{N_c+1}$ and the
rest having $R$-charge $\frac{2}{N_c+1}$ corresponds to the IR $s$-confining
phase of SU($N_c$) SQCD with $N_c+1$ flavors, where the chiral matter described
above are the confined mesons and baryons. In particular, for $N_c=2$ the mesons and baryons attain the expected superconformal $R$-charges for a free theory.

The index is
\begin{eqnarray}
\mathcal{I}_{s.t.}&=&N\mathcal{I}_{s.t.}(V)+\sum_{i=1}^{N_{\chi}}\mathcal{I}_{s.t.}(\chi,\bar{\chi},R_i) \nn \\
&=& N\left(1-\frac{1-t^{-2}}{(1-t^{-1} y)(1-t^{-1}
y^{-1})}\right)+\sum_{i=1}^{N_{\chi}}\frac{t^{-R_i}-t^{R_i-2}}{(1-t^{-1}
y)(1-t^{-1} y^{-1})},\label{eq:U(1)Nindex}
\end{eqnarray}
which when combined with Eq.~(\ref{eq:c-aIndex}) gives
\begin{eqnarray}\label{eq:U(1)Nc-a}
c-a&=&\left[-\frac{N}{16}- \sum_{i=1}^{N_{\chi}}\frac{
R_i-1}{16}+\mathcal{O}(t-1)\right]^{\mbox{\scriptsize{finite}}}_{y=1,t\rightarrow
1} \nn\\
&=&-\frac{N}{16}- \sum_{i=1}^{N_{\chi}}\frac{R_i-1}{16},
\end{eqnarray}
which is indeed $-\fft{1}{16}\mathrm{Tr}R$. Curiously, this
expression for $c-a$ is finite as $t\to1$.

\subsubsection{Large-N SQCD}

We now consider standard SQCD \cite{Seiberg:1994pq}. This theory has
a large-$N$ limit which does not admit a dual gravity description.
In particular, $c-a$ for this theory is \cite{Kutasov:1995np}
\begin{equation}
c-a = \frac{1}{16}(N_c^2+1)
\end{equation}
which is $\mathcal O(N^2).$ Nevertheless, we will see that our
prescription successfully computes $c-a$ from the large-$N$
expression of the index.

Consider SQCD in the Veneziano limit $N_c, N_f \gg 1$ with $N_c/N_f$
held fixed. The superconformal index for this theory has been
computed in \cite{Dolan:2008}, and the single-trace index can be obtained by
taking its plethystic log.  After simplification, the result is given (in our notation) by
\begin{equation}
\mathcal I_{s.t.} = \frac{1}{t^2-1}\left(1 + N_f^2
\frac{(t^{N_c/N_f}-t^{-N_c/N_f})^2}{(1-t^{-1}y)(1-t^{-1}y^{-1})}\right).
\end{equation}

Applying Eq.~(\ref{eq:c-aIndex}) to this we find
\begin{equation}
c-a = \left[-\frac{3}{16(t-1)^2} + \frac{1}{16}(N_c^2+1) +
\mathcal{O}(t-1)\right]^{\mbox{\scriptsize{finite}}}_{t\rightarrow
1}=  \frac{1}{16}(N_c^2+1),
\end{equation}
which recovers the expected value of $c-a.$

\subsubsection{$A_k$ Theories}

Next, consider the $A_k$ generalizations of SQCD
\cite{Kutasov:1995np} which add an additional adjoint chiral superfield $X$
along with a superpotential of the form
\begin{equation}
W = \tr X^{k+1}.
\end{equation}
For $k=1$ this is just a massive deformation, and the $X$ field can
be integrated out to yield the standard SQCD of the previous
subsection. For generic $k$ the difference of central charges is
given by
\begin{equation}
c-a = \frac{1}{8(k+1)}(N_c^2 +1).
\end{equation}
\begin{widetext}
The superconformal index has been computed for these theories in the
Veneziano limit \cite{Dolan:2008}, and we can again extract the
single-trace index to obtain
\begin{equation}
\mathcal I_{s.t.} = \frac{t^{-\frac{2}{k+1}}}{1-t^{-\frac{2}{k+1}}}
+ \frac{t^{-\frac{4k}{k+1}}}{1-t^{-\frac{4k}{k+1}}} -
\frac{t^{-\frac{2k}{k+1}}}{1-t^{-\frac{2k}{k+1}}} - \frac{
\big(t^{-\frac{2}{k+1}} - t^{-\frac{2k}{k+1}}\big) - N_f^2
\frac{\big(t^{\frac{2N_c}{(k+1)N_f}} -
t^{-\frac{2N_c}{(k+1)N_f}}\big)^2}{t^2\big(1-t^{-\frac{2}{k+1}}\big)\big(1+t^{-\frac{2k}{k+1}}\big)}}{(1-t^{-1}y)(1-t^{-1}y^{-1})}.
\end{equation}
From this we find
\begin{equation}
c-a = \left[-\frac{3(2k-1)(k+1)}{32k(t-1)^2} +
\frac{1}{8(k+1)}(N_c^2 +1) +
\mathcal{O}(t-1)\right]^{\mbox{\scriptsize{finite}}}_{t\rightarrow1}
= \frac{1}{8(k+1)}(N_c^2 +1),
\end{equation}
\end{widetext}
and we again arrive at the expected value for $c-a.$

\section{A Comment on $\mathcal N=2$ to $\mathcal N=1$ RG Flows}

For theories preserving $\mathcal N=2$ supersymmetry the index can
be further refined by introducing a fugacity for the enlarged
$SU(2)_R\times U(1)_r$ $R$-symmetry. In this case the right-handed
index can be written (in this section we are explicitly setting
$a_i=1$ for all $i$)
\begin{equation}
\mathcal I^R_{\mathcal N = 2} = \tr
(-1)^{F}e^{-\beta\delta}t^{-2(E+j_{2})/3} y^{2j_{1}}v^{-(r_{\mathcal
N=2} + R_{\mathcal N=2})},
\end{equation}
where $R_{\mathcal N=2}$ and $r_{\mathcal N=2}$ are the quantum numbers under the $SU(2)_R\times U(1)_r$ $R$-symmetry which are related to those of the $\mathcal N=1$ $R$-symmetry by
\begin{equation}
r_{\mathcal N=1} = \frac{2}{3}(2R_{\mathcal N=2} - r_{\mathcal N=2}).
\end{equation}

In general, $\mathcal N=2$ theories can be deformed by giving a mass
term to the $\mathcal N=1$ chiral multiplet that sits in the
$\mathcal N=2$ vector multiplet. In this case, one arrives at an
$\mathcal N=1$ theory in the infrared (IR). When all of the gauge
couplings are exactly marginal and when one gives mass terms to the
chiral multiplets in all of the $\mathcal N=2$ vector multiplets,
Ref.~\cite{Gadde:2010en} demonstrated an interesting relation
between the $\mathcal N=2$ index of the ultraviolet (UV) theory and
the $\mathcal N=1$ index of the IR theory of such RG flows. In
particular, they showed that the IR $\mathcal N=1$ index is given by
simply setting the fugacity $v = t^{-1/3}$ in the UV $\mathcal N=2$
index, so that
\begin{equation}\label{eq:N2N1index}
\mathcal I^{\text{UV}}_{\mathcal N=2}(t,y,v=t^{-1/3}) = \mathcal
I^{\text{IR}}_{\mathcal N=1}(t,y).
\end{equation}
This implies a simple relation between the flow of $c-a$,
parametrized by $\Delta_{c-a} \equiv (c-a)_{\text{UV}} -
(c-a)_{\text{IR}}$, and the $\mathcal N=2$ index of the theory in
the UV. There are known universal relations between the UV and IR
central charges of such theories \cite{Tachikawa:2009tt}, and it
would be interesting to know if these results are related to what we
discuss below.

Notice first that one can evaluate the $\mathcal N=1$ index for an
$\mathcal N=2$ theory by simply setting $v=1$, and so our
prescription for computing $c-a$ from the index applies also to
theories preserving more supersymmetry with the additional
fugacities set to unity.  We now apply the observation
(\ref{eq:N2N1index}) to the expression for $c-a$ and use it to
compute $\Delta_{c-a}$. The difference is given by
\begin{widetext}
\begin{eqnarray}\label{eq:diffc-a}
\Delta_{c-a} &=&  \lim_{t\to1}-\frac{1}{32}\left(t\partial_t
+1\right)\left(6(y\partial_y)^2-1\right)\nn\\
&&\times\left[(1-t^{-1} y)(1-t^{-1} y^{-1})\left(\mathcal
I^{\mathcal N=2}_{s.t.}(t,y,v=1)-\mathcal I^{\mathcal
N=2}_{s.t.}(t,y,v=t^{-1/3})\right)\right]^{\mbox{\scriptsize{finite}}}_{y=1}.
\end{eqnarray}
When acting on the second term in the parentheses, the $t$
derivative can be re-written as
\begin{equation}
t\partial_t+t\frac{dv}{dt}\partial_v = t\partial_t -
\frac{1}{3}v\partial_v,
\end{equation}
where one should afterwards set $v=t^{-1/3}$ before taking the limit
in (\ref{eq:diffc-a}).

If the index expressions were finite, then the only terms that
survive in the difference once we take $t\to1$ are those in which
the $v$ derivative acts explicitly on the IR index. In this case,
the result can then be written purely in terms of the $\mathcal N=2$
index
\begin{equation}\label{eq:deltac-a}
\Delta_{c-a} =
\lim_{t\to1}-\frac{1}{96}v\partial_v\left(6(y\partial_y)^2-1\right)\left[(1-t^{-1}
y)(1-t^{-1} y^{-1})\mathcal I^{\mathcal
N=2}_{s.t.}(t,y,v)\right]\Big|^{\mbox{\scriptsize{finite}}}_{y=1,\,v=t^{-1/3}}.
\end{equation}
In fact, this is only strictly true for functions which are finite
as $t$ approaches one, as would be the case when (\ref{eq:diffc-a})
is applied to the single-trace index at finite $N$. However we know
the result to be divergent at $t=1$ in the large-$N$ limit from the
examples studied in the previous sections. Nevertheless, since we
expect (\ref{eq:deltac-a}) to hold for the index at finite $N$, we
should expect it also to be true in the large-$N$ limit. This can
be seen in a simple example.

Consider the Klebanov-Witten flow between the holographic duals to
the $S^5/\mathbb Z_2$ orbifold theory and the $T^{1,1}$ theory
\cite{Klebanov:1998hh}. The single-trace $\mathcal N=2$ index for
the $S^5/\mathbb Z_2$ theory is \cite{Gadde:2009dj}
\begin{equation}
\mathcal I^{\mathcal N=2}_{s.t.} =
2\left(\frac{t^{-2/3}v}{1-t^{-2/3}v} +
\frac{t^{-4/3}/v}{1-t^{-4/3}/v} -
\frac{t^{-2/3}v-t^{-4/3}/v}{(1-t^{-1}y)(1-t^{-1}y^{-1})} \right).
\end{equation}
Note that setting $v=1$ reproduces the $\mathcal N=1$ index given in (\ref{eq:Z2orbi}).  We can
evaluate (\ref{eq:deltac-a}) for this theory to find
\begin{equation}
\Delta_{c-a}(S^5/\mathbb Z_2\rightarrow T^{1,1}) = - \frac{1}{24},
\end{equation}
which is the expected result.
\end{widetext}

It is also worth noting that this result is completely finite at
$t=1.$ Also, if one evaluates the remaining terms in
(\ref{eq:diffc-a}), {\it i.e.}\ those that would vanish for a
function that is regular at $t=1,$ we find that they correctly
reproduce the $(t-1)^{-2}$ divergence of the difference of the $c-a$
results from applying (\ref{eq:c-aIndex}) directly. In particular,
they contain no finite term. It would be interesting to understand
if $\Delta_{c-a}$ as defined by (\ref{eq:deltac-a}) is finite
generically.

\section{Moving away from the large-$N$ limit}

Since the relation (\ref{eq:c-aIndex}) was motivated by holographic
considerations, it is natural for it to hold in the large-$N$ limit.
However, one can ask whether it will remain valid even at finite $N$
(provided the single-trace index is replaced by the plethystic log
of the full index).  One possible obstruction in making this
connection between large and finite $N$ arises from the
regularization of the $t\to1$ divergence that is present at large
$N$.  As seen in the above examples, this divergence shows up as a
second order pole in $t-1$, which needs to be removed by hand in
order to obtain a finite answer for $c-a$. (One exception to this
behavior is in the $U(1)^N$ theory of
Section.~\ref{subsubsec:U(1)N}, for which $N$ is not necessarily
large.)

Because of the $t$-derivative in (\ref{eq:c-aIndex}), we see that
the second order pole in $c-a$ arises from a first order pole in the
single-trace index. Pole terms in the index arise due to geometric
sums over an infinite number of states. The reason the $U(1)^N$
theory avoids this pole is that the only infinite series of states
at finite-$N$ is due to the descendent operators. In this case the
single-trace index multiplied by $(1-t^{-1}y)(1-t^{-1}y^{-1})$ (to
remove the descendent contributions) is finite at $t=1$ and
fixed $y\neq 1$ for finite $N$.
This may be understood physically by realizing that only a finite
number of protected single-trace operators may arise in the abelian
theory.  More generally, the number of such operators in a gauge
theory with gauge group SU($N$) is of order $N$, since traces of
products with more than $N$ terms can be written in terms of
products of shorter ones.  So a theory at finite-$N$ implicitly cuts
off the potential geometric sums over single-trace states. What this
suggests is that the pole term of $c-a$ in $t-1$ arises as an
artifact of the large-$N$ limit.

The combination of $t$ and $y$ serve to regulate the
index, so the $t\to1$ and $y\to1$ limits are somewhat delicate. When
the descendant states are removed from the single-trace index by
multiplication with $(1-t^{-1}y)(1-t^{-1}y^{-1})$, the expression at
$t=1$ essentially counts the number of protected single-trace
operators in the theory.  In this case it ought to remain finite at
$t=1$ when $N$ is finite. On the other hand, the single-trace index
by itself at finite $N$ diverges when $y=1$ and $t\to1$ because of
the contribution of the descendant states. Considering again the
$U(1)^N$ example, we see that the finite-$N$ single-trace index at
$y=1$ has a first order pole at $t=1$ with structure given by
\begin{equation}
\mathcal I^{\mathrm{finite\mhyphen}N}_{s.t.}(t\rightarrow1,y=1)
\simeq -2\frac{\mathrm{Tr}R}{t-1}=32\frac{c-a}{t-1}.
\label{eq:finiteTrR}
\end{equation}
This behavior of the index has also been shown to be generically
true \cite{DiPietro:2014bca}. In particular, we can see this by defining
$t = e^\beta$ and taking the Plethystic exponential of the above
single-trace index. Doing this for this example and extracting the
leading $\beta \rightarrow 0$ behavior gives
\begin{widetext}
\begin{equation}
\mathcal I^{\mathrm{finite\mhyphen}N} = \exp\left(\sum_{n=1}^\infty
\frac{1}{n}\mathcal I_{s.t.}^{\mathrm{finite\mhyphen}N}(e^{n\beta},y^n)
\right)\Big|_{y=1} \simeq \exp\left(\frac{32(c-a)}{\beta}\sum
\frac{1}{n^2}\right) = \exp\left(\frac{16\pi^2 (c-a)}{3 \beta}\right),
\end{equation}
which precisely agrees with the general results of \cite{DiPietro:2014bca}.

For large-$N$ single-trace indices, even when $y\neq1$ there are
pole terms around $t=1$ because at infinite $N$ there are infinitely
many protected single-trace operators in the theory. We have not
been able to recognize the coefficient of this pole in general.
However, it is instructive to examine the pole structure somewhat
more carefully.

The behavior of $\mathcal{I}_{s.t.}$ as $t\to1$ and $y\to1$ depends
on how the limit is taken.  By taking $t\to1$ first, we obtain an
expansion of the form
\begin{equation}
\mathcal{I}_{s.t.}(t,y)=\frac{a' _0}{t-1}+b' _0+(t-1)
\left(\frac{c'_{-2}}{(y-1)^2}+\frac{c'
_{-1}}{y-1}+c'_0\right)+\cdots.\label{eq:tauFirst}
\end{equation}
Although we do not claim that this is the most general structure, it
nevertheless holds for all examples we have looked at.  In this
expansion, there is at most a simple pole in $t-1$, while the higher
order terms have increasing poles in $y-1$.  In contrast, taking
$y\to1$ first yields the expansion
\begin{eqnarray}
\mathcal{I}_{s.t.}(t,y)&=&\frac{a _0}{t-1}+a_1+a_2 (t-1)+\cdots\nn\\
&&+(y-1)^2\left(\fft{b_0}{(t-1)^3}+\fft{b_1}{(t-1)^2}+\fft{b_2}
{t-1}+b_3+b_4(t-1)+\cdots\right)+\cdots.\label{eq:yFirst}
\end{eqnarray}
\end{widetext}
That there is no term proportional to $(y-1)$ in the above expansion
follows from CP invariance.  While we do not have an argument to
demonstrate that the leading pole in the second line is third order
in $t-1$, this nevertheless holds for all of the examples we have
considered.

More generally, it is possible to approach $t\to1$ and $y\to1$ from
different directions.  For example, if we considered the supersymmetric
partition function on the squashed three-sphere times a circle \cite{Imamura:2012},
then the natural limit would be to hold the squashing parameter $b$ fixed.
The approach to $t\to1$ and $y\to1$ then follows the curve
$y=t^{(1-b^2)/(1+b^2)}$.  (Note that this reduces to $y=1$ for the
round $S^3$.)  This choice of holding $b$ fixed was taken in \cite{DiPietro:2014bca}.

A point to be made about the two expansions in (\ref{eq:tauFirst})
and (\ref{eq:yFirst}) is that the coefficients $a_0'$ and $a_0$ of
the pole terms do not necessarily agree, as demonstrated in
Table~\ref{tbl:Icoefs}, because the order of the limits $y\to1$ and
$t\to1$ is important. The expansion that is relevant for computing
$c-a$ using (\ref{eq:c-aIndex}) is (\ref{eq:yFirst}), where the
expansion around $y=1$ is only needed to second order. In this case,
we obtain
\begin{equation}
c-a=-\frac38\frac{a_0-b_0}{(t-1)^2} +\frac{1}{32}(a_0
+12(a_2-b_0+b_1-b_2))+\cdots.\label{eq:c-ayFirst}
\end{equation}
Moreover, the vanishing of the simple pole term in $c-a$ follows directly from the
form of the operator $t\partial_t+1$ in (\ref{eq:c-aIndex}) acting on (\ref{eq:yFirst}).
(This would not be true if we had chosen a different convention for the normalization
of $t$.)  It is the finite part in (\ref{eq:c-ayFirst}) that correctly reproduces the value
expected from weak-coupling field theory computation, namely
$-\fft1{16}\mathrm{Tr}R$, in all examples we have considered. Note in
particular that for the large-$N$ Veneziano limit of $A_k$ theories,
$c-a=(N_c^2 +1)/8(k+1)$ is large (in contrast with the holographic
examples), but Eq.~(\ref{eq:c-aIndex}) still works perfectly fine.
The coefficients relevant to $c-a$ are listed for a number of
examples in Table~\ref{tbl:Icoefs}.

\begin{table*}[tp]
\centering
\begin{tabular}{|l|l|l|l|l|l|l|l|}
\hline
Theory &$a' _0$&$a_0$&$a_1$&$a_2$&$b _0$&$b_1$&$b_2$\\
\hline
Toric w/o adjoint&$3\sum \frac{1}{r_i}$&$3\sum \frac{1}{r_i}$&$-\ft{n_z}{2}+\frac{3}{2}\sum\frac{1}{r_i}$&$\frac{1}{36}\sum r_i-\frac{1}{4}\sum\frac{1}{r_i}$&$0$&$0$&$0$\\
$\mathcal{N}=4$ theory&$9/2$&$5/2$&$-1/4$&$19/216$&$-2$&$-3$&$-19/27$\\
$\mathbb{Z}_2$ orbifold theory&$9/2$&$19/6$&$-5/12$&$101/648$&$-4/3$&$-2$&$-38/81$\\
SPP quiver&$9$&$\frac{13}{2}-\frac{4}{\sqrt{3}}$&$\frac{3}{4}-\frac{2}{\sqrt3}$&$\frac{35}{24}-\frac{20}{9\sqrt{3}}$&$2-\frac{4}{\sqrt{3}}$&$3-2\sqrt3$&$\frac{7}{3}-\frac{38}{9\sqrt3}$\\
SQCD&$1/2$&$1/2+2N_c^2$&$-1/4+N_c^2$&$\fft18-\fft{2N_c^2}3+\fft{2N_c^4}{3N_f^2}$&$2N_c^2$&$3N_c^2$&$\fft{N_c^2}3+\fft{2N_c^4}{3N_f^2}$\\
\hline
\end{tabular}
\caption{\label{tbl:Icoefs} The coefficients of the expansions
(\ref{eq:tauFirst}) and (\ref{eq:yFirst}) of the single-trace index
at large $N$.}
\end{table*}

As seen in (\ref{eq:c-ayFirst}), the regulated $c-a$ receives
contributions from several terms in the expansion of the
single-trace index.  In particular, it includes information away
from $y=1$, as encoded in the $b_0$, $b_1$ and $b_2$ coefficients.
This is, of course, due to the $y$-dependent operator in
(\ref{eq:c-aIndex}), which is needed to bring in the proper
spin-dependence to reproduce the holographic expressions
(\ref{eq:CHc-a}) and (\ref{eq:sl2c-a}).  In this sense, it appears
that while the information of $c-a$ is contained in the large-$N$
index, it is encoded in a rather non-trivial manner.

The expansion (\ref{eq:c-ayFirst}) also applies to the $U(1)^N$
theory of Section.~\ref{subsubsec:U(1)N}. For this particular case
it turns out that $a_0=b_0$ so that the pole term in $c-a$ vanishes.
Furthermore, while all the coefficients in the expansion
(\ref{eq:yFirst}) are non-vanishing, it turns out that the sum
$(a_2-b_0+b_1-b_2)$ also vanishes. This leaves only the finite
$a_0/32$ as the result for $c-a$, which agrees with
(\ref{eq:finiteTrR}), {\it i.e.}\ that the coefficient of the pole
term of the finite-$N$ single-trace index is proportional to $c-a.$%
\footnote{We would like to thank Z.\ Komargodski for discussions on
this point.}

In fact, the observation of the preceding paragraph can be generalized
assuming a particular form for the index. Let us suppose that the finite-$N$
index has a pole structure determined only by the descendent contributions such
that it has the form
\begin{equation}
\mathcal I^{\text{finite-}N}_{s.t.}(t,y) =
\frac{F(t,y)}{(1-t^{-1}y)(1-t^{-1}y^{-1})},\label{eq:finiteNguess}
\end{equation}
where $F(t,y)$ is a regular function with a first order zero at
$t=1$ when $y=1$. This behavior of the function $F(t,y)$ can be seen heuristically by considering the fact that in an expansion around $t=y=1$ one expects the bare sum (i.e. neglecting the denominators from the descendent contributions) on single trace operators to vanish since the index ought to pick up a contribution from an equal number of bosonic and fermionic states, thus leading to (at least) a first order zero. This can be seen in $\mathcal N \ge 2$ theories because states contributing to the index live in representations of a supersymmetric commuting sub-group of the superconformal group. For $\mathcal N=1$ the commuting sub-group no longer contains a fermionic generator. Nonetheless one can see a cancelation of this sort by grouping all chiral (or semi-long) operators into triplets which, together with the bare chiral (semi-long) operator, correspond to inserting e.g. the gaugino superfields $W_\alpha$ and $W^\alpha W_\alpha$ into the trace.  The sum of contributions to the single-trace index over any such triplet indeed vanishes at $t=y=1.$ (Note that this argument breaks down if one must sum over an infinite tower of states, as we have seen in the large-$N$ examples discussed previously, in which case $F(t,y)$ acquires a pole term in $t$ for $y$ not strictly equal to unity. This leads to the second order pole in $c-a.$) In this case, one can Taylor expand the numerator
near $t=y=1$
\begin{eqnarray}
F(t,y) &=& f_1(t-1) + f_2(t-1)^2 + f_3(t-1)^3 + \cdots\nn\\
&& + (y-1)^2 \big(g_0 + g_1(t-1) + \cdots\big) + \cdots,
\end{eqnarray}
where we have kept only the terms relevant for the calculation of $c-a$ in (\ref{eq:c-ayFirst}).
Assuming the above form and computing $c-a$ via (\ref{eq:c-ayFirst}), one finds again that the
coefficients in the expansion (\ref{eq:yFirst}) satisfy $a_0=b_0$. However, the sum
$(a_2-b_0+b_1-b_2) = - g_0 - g_1.$ While we do not have an argument for the vanishing
of the linear combination of coefficients $g_0 + g_1$, if it does vanish, then we
recover the result that $c-a$ is simply proportional to the coefficient $a_0$
of the pole term in the expansion (\ref{eq:yFirst}). It would be interesting to see if a general
argument exists that implies the structure assumed in (\ref{eq:finiteNguess}) and that
also enforces $g_0 + g_1 = 0.$

A more heuristic argument for the vanishing of the pole term in
(\ref{eq:c-ayFirst}) for finite $N$ theories is that in this case it
appears that only a finite number of protected operators would enter
into the computation of $c-a$.  This would require the plethystic
log of the index (which we take to be the finite-$N$ counterpart of
the single-trace index) to satisfy the condition $a_0=b_0$ (while
$a_0$ itself is non-zero), regardless of the regularity of $F(t,y)$
in (\ref{eq:finiteNguess}).  It is not clear to us what the
significance of this relation is, and how it sees the distinction
between finite $N$, where $a_0=b_0$ may plausibly hold, and large
$N$, where it is clearly violated.  A better understanding of the
structure of the finite-$N$ index may be needed in order to clarify
this condition and to explore the applicability of
(\ref{eq:c-aIndex}) at finite $N$.

As indicated above, the pole term in $c-a$ is non-vanishing in the large-$N$ limit.  This is essentially
the counterpart of the divergence in the corresponding holographic computation
\cite{Ardehali:2013gra,Ardehali:2013xla,Ardehali:2013xya}.  In particular, the index provides
a natural regulator of the sum over the shortened spectrum.  For theories
admitting a holographic dual, we conjecture (guided by
\cite{Eager:2010dk} and following \cite{Ardehali:2013gra}) that the pole term may be related to
geometric data given by
\begin{equation}
-\frac{3(a_0-b_0)}{8}=-\frac{9}{128\pi
^3}\left(19\,\mbox{vol}(\mbox{SE})+\frac{1}{8}\mbox{Riem}^2(\mbox{SE})\right).
\label{eq:pole}
\end{equation}
This is valid for all the holographic theories dual to smooth SE$_5$
manifolds that we considered in the present work.

It would be interesting to apply (\ref{eq:c-aIndex}) to the finite-$N$ theories
where the full index is known as a means of testing the conjecture that this relation remains
valid at finite $N$.  A possible issue is resolving how single- versus multi-trace operators
contribute to $c-a$, and whether such contributions are properly handled by the plethystic log.
Ideally, one would wish to replace (\ref{eq:c-aIndex}) by some sort of operator expression
acting directly on the index without having to resort to the plethystic log.  Ref.~\cite{DiPietro:2014bca}
suggests that a simple relation exists of the form (\ref{eq:finiteTrR}).  It would be curious to see
how this connects with (\ref{eq:c-aIndex}), both at finite and large $N$.

Finally, at least in the large-$N$ limit, we have demonstrated that
the superconformal index contains information about the difference
$c-a$.  At some level, this is perhaps not all that surprising, as
the holographic computation of $c-a$ depends only on the spectrum of
protected operators, and the index contains the full information
about such operators.  One may ask whether it is possible to obtain
$a$ and $c$ individually from the index.  However, the holographic
computation suggests that this would be a challenge, as long
operators (which are not captured by the index) will contribute to
the individual central charges.  It may nevertheless be possible
that the protected operators themselves retain sufficient
information to allow for the extraction of $a$ and $c$ separately
(see \cite{Razamat:2012,Kim:2012ava,Buican:2014,Assel:2014} for
interesting related discussions)\footnote{See in particular
appendix~B of \cite{Kim:2012ava} for relations between the central
charges and the \emph{single-letter} index.}. In any case, it is
clear that the index encodes a wide range of information that can be
used to more fully characterize the SCFT, and is an important tool
for their investigation, especially at strong coupling.

\begin{acknowledgments}
We thank I.~Bah, L.~Di Pietro, R.~Eager, H.~Elvang, Z.~Komargodski,
D.~Martelli, Y.~Tachikawa, and B.~Wecht for interesting
conversations about this project. We are grateful to L.~Di Pietro
and Z.~Komargodski for sharing with us an advance copy of
\cite{DiPietro:2014bca} and to R.~Eager for guiding us to the
derivation of Eq.~(\ref{eq:sumRule}). This work was supported in
part by the US Department of Energy under grants DE-SC0007859 and
DE-SC0007984.
\end{acknowledgments}

\end{document}